Pierre Barre, Chaouki Kasmi and Eiman Al Shehhi

Mobile and Telecom Lab,

Xen1thLabs, a Dark Matter Company

Abu Dhabi, United Arab Emirates

# Spy the little Spies

*Security and Privacy issues of Smart GPS trackers*

**Abstract**-Tracking expensive goods and/or targeted individuals with high-tech devices has been of high interest for the last 30 years. More recently, other use cases such as parents tracking their children have become popular. One primary functionality of these devices has been the collection of GPS coordinates of the location of the trackers, and to send these to remote servers through a cellular modem and a SIM card. Reviewing existing devices, it has been observed that beyond simple GPS trackers many devices intend to enclose additional features such as microphones, cameras, or Wi-Fi interfaces enabling advanced spying activities. In this study, we propose to describe the methodology applied to evaluate the security level of GPS trackers with different capabilities. Several security flaws have been discovered during our security assessment highlighting the need of a proper hardening of these devices when used in critical environments.

**Disclaimer**-The opinions and results presented in this article are the sole responsibility of the authors.

## I. Introduction

Sometimes close to illegal activities in certain countries, the deployment of GPS trackers to follow individuals and expensive commercial goods has regained interest in recent years. With the main purpose of sending GPS coordinates of the location of tracked devices or individuals, the GPS trackers are automatically and periodically collecting data and sending these to remote servers through cellular networks by using data connection or text messages.

While reviewing multiple tracker solutions, it has been observed that the manufacturers of these devices have added smart features making these devices more than simple GPS tags, but fully operational spying devices with movement/noise detectors, microphones, cameras and additional wireless interfaces.

The configuration of such devices is implemented through proprietary protocols over text messages. Before deploying such devices, users will have to configure a set of parameters such as the remote server IP address and the identification and authentication credentials, in order to, for examples, uploading the collected data to the cloud infrastructures.

From a security/privacy perspective it is obvious that an attacker who is able to get access to the phone numbers of the SIM cards provisioned in these trackers may endanger the covertness of operations or/and the expensive goods location use this information to mount targeted attacks.

Only a few studies have been released covering the entire attack surface exposed by these devices. For instance, the valuable Trackmagedon [1] study highlights the security flaws existing in multiple cloud management system where the collected data originates from GPS trackers. Nevertheless, when looking at the architecture and protocols of such devices when deployed on the field, it can be imagined that an attacker would not only be interested in the cloud management infrastructures, but also targeting the GPS tracking devices through the mobile network (e.g. SMS configuration interface) to get access to the advanced features like the microphone or the stop engine (feature found in some GPS trackers to provide remote access to the car engine) feature. In this study, we propose to analyze the possible attack scenarios and attack vectors exposed by GPS trackers available on the market. From zero to full take-over of the infrastructure, a detailed security assessment is proposed, highlighting security critical vulnerabilities that are putting the users of such solutions at risk.

The study is composed as follows: in Section II, we propose a review of the attack surface and potential attack vectors exposed by GPS trackers. Then, in Section III, a security assessment regarding the described attack vectors is summarized. Finally, conclusions are drawn highlighting the criticality of identified vulnerabilities and their potential impact on targeted people and exposed goods in terms of security and privacy. Moreover, the collection of data by these devices is discussed from a national security perspective.

## II. Attack Vectors

A diagram representing the general architecture and points of interaction is proposed in Figure 1. As it can be observed, the attack surface is fairly large considering the amount of possible radio interfaces (GPS, mobile networks), the remote servers and running web applications, the smartphone applications and the configuration management protocols. Moreover, updating mechanisms (if they exist) are also interesting attack vectors to exploit. However, the set of tested devices in this study did not implement remote update mechanisms.

In this section, we describe possible entry points for an attacker according to his objective. Note that some of these attacks may be illegal in certain countries. It is further worth to mention that all techniques and tools used in this study are known and available online either open source or as free versions of the software.

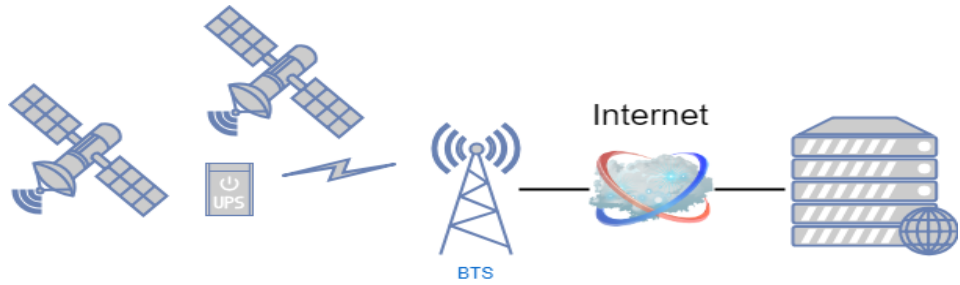

*Figure 1: Architecture of the infrastructure related to GPS trackers with cellular modem*

## II.A. Jamming and Spoofing attacks against the GPS signal

An attacker whose goals are to either report wrong GPS coordinates or to block the transmission to remote servers can use low-cost GPS jammers or GPS spoofing tools. Low-cost GPS jammers [2] can be found online in different formats. These attacks have already been detected across the world as reported in "*140,000 unique electronic signatures for GPS jammers in Europe*" [3] in 2017, highlighting the need of proper hardening mechanism for GPS trackers.

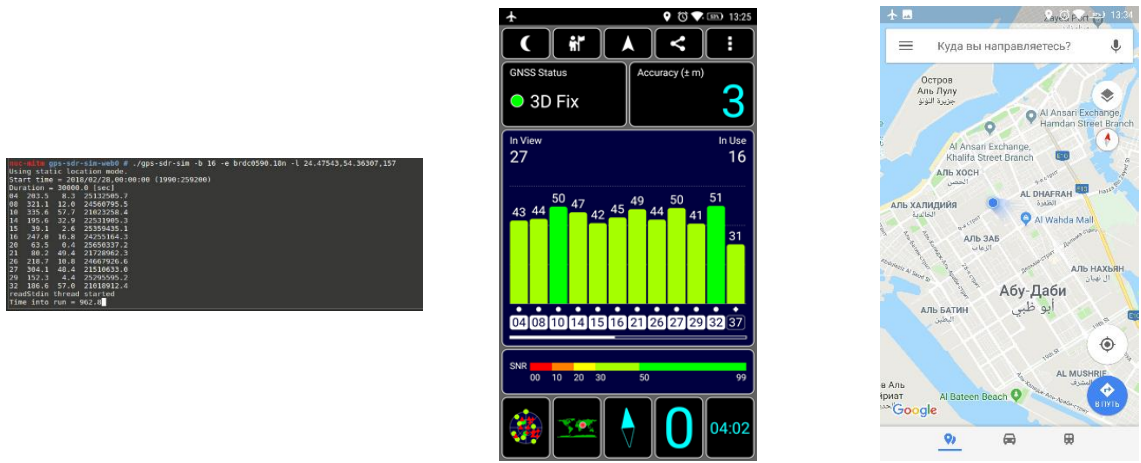

*Figure 2: Spoofing attack performed against UAVs, Smartphone and GPS trackers*

Advanced GPS attacks, namely GPS Spoofing, have been publicly described [4] and open source tools have been released online. Thus, using a software defined radio and the *gps_sdr_sim* software [5], an attacker is able to emulate a specific GPS satellite constellation in the vicinity of the targeted devices. The tools have been used for different sets of targets as depicted Figure 2 such as smartphones and drones. Note that an easy countermeasure would be to check third party information such as Cell_IDs to cross-validate the location of a specific tracker.

Spy the little Spies                                                                                    3

**II.B Attacks against the management protocols**

With the goal of reconfiguring trackers to get access to the GPS coordinates in real-time, an attacker can try to change the IP addresses of the management servers and force trackers to transmit GPS coordinates to its own infrastructure. In order to avoid being detected, he can then transmit the data after modifying them on the fly towards the legitimate infrastructure. This attack obviously requires knowing the management protocol and the phone number of the SIM cards provisioned in trackers to exploit potential vulnerabilities in the management protocol. The complexity of this attack is directly related to the level of security of the management protocol and its implementation. Interestingly, plenty of GPS trackers user manuals are available online as for example on the FCC website [6].

**II.C. Attacks against the cloud management, web application and smartphone applications**

An attacker may target the infrastructure to compromise the entire dataset of GPS trackers for a given manufacturer as demonstrated to other type of IoTs such as IP cameras [7]. The objective would be to directly attack the remote management infrastructure and the web/mobile applications available. By obtaining a set of trackers, the attacker will be able to reverse engineer protocols, the firmware, and the hardware parts. Then, an analysis of the mechanisms of authentication, identification, and available APIs will be leveraged to find security flaws leading to a point of entry. The level of accessible information makes this attack scenario the most effective since it potentially allows attacking a large number of trackers simultaneously.

In what follows, we propose to go through different experimental set-ups and to summarize a set of security flaws demonstrating the low level of security of GPS trackers and their associated infrastructures. In order to make this study as generic as possible, we have procured the best-selling GPS trackers online.

**III. Security evaluations**

The methodology applied to analyze the GPS trackers and their infrastructures is well known by security researchers. Starting by the analysis of each component (hardware, software and network), the security assessment can then be tailored to specific scenarios.

In this section, we summarize security flaws we discovered regarding different attack scenarios and possible exploitation techniques (when not fully performed in this study). Moreover, the complexity and the criticality of these issues are discussed.

### III.A. Data collecting and forwarding

As a first step, a network traffic analysis was carried out to understand the type of data exchanged between the GPS trackers and the remote servers as well as the communication protocols. In a black-box context, we set up a base-station 2G/2.5G (GPRS) with a generic configuration of the Yatebts [8]. It is noted that the majority of trackers only support SIM cards due to the integrated baseband.

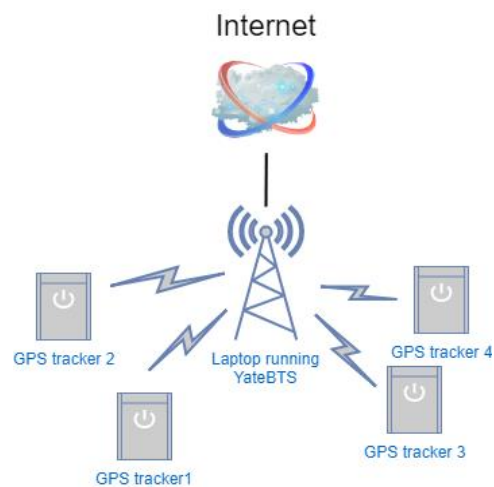

*Figure 3: Setup for capturing data sent from GPS trackers to the remote servers*

For regulatory constraints, the set of GPS trackers has been placed in a Faraday cage. Each of our GPS trackers is equipped with a SIM card and an assigned unique number as represented on Figure 3. Once the BTS was operational, the GPRS traffic was captured with the *sgsntun* interface. Using the provided user manual, the SMS commands recommended by the manufacturer to parameterize each of our trackers have been send. Interestingly, during this configuration phase, the following critical items have been observed:

- all trackers send coordinates to IPs in China – confirmation of the location of remote servers;
- the traffic is not encrypted and is easily identifiable (specific domain names, specific IPs, specific ports) - the implementation deep packet inspection tool is trivial to perform for this type of protocol;

- the password of one platform was sent in plain text, other platforms use the serial number of the trackers (predictable) as a token without authentication mechanism, allowing an attacker to send forged information;
- it is possible, via an SMS, to edit the tracker configuration and specify another management server ("`* reg my_ip`"), allowing an attacker to perform a simple MITM via an Internet server relaying the traffic (UDP or TCP - according to the trackers), with for example `balance (1)` for TCP:

    ```
    balance -b ::ffff:mon_ip 8841 203.130.62.29:8841
    ```

- it is possible to define a master phone number. In case of Caller-ID spoofing, this authentication can be bypassed. In addition, even without having the defined master phone number, it is possible to send commands via SMS to retrieve information from the GPS tracker, thus making it possible to detect trackers for a range of phone numbers - this last method will not be discrete.
- Trackers using a U-BLOX GPS module connect to a TCP service on the 56447/tcp port. By default, the client sends his login, password, latitude, longitude and altitude without integrity or confidentiality protection at the beginning of the TCP session:

    ```
    cmd=full;user=XXXXXX@gmail.com;pwd=XXXXXX;lat=22.680193;lon=114.146846;alt=0.0;pacc=100.00
    ```

The server responds by specifying a proprietary blob:

```
u-blox a-gps server (c) 1997-2009 u-blox AG
Content-Length: 2856
Content-Type: application/ubx
.b..0......
```

The client then regularly sends information to a second server (8011 / tcp) indicating its position:

```
*HQ,17000XXXXX,V1,115112,A,2240.8116,N,11408.8108,E,000.0,000.00,100119,FFFFFFFF#

*HQ,17000XXXXX,NBR,094111,310,26,02,1,1000,10,23,100119,FFFFFFFF#
```

```
*HQ,17000XXXXX,LINK,115112,22,0,6,0,0,100119,FFFFFFFF#

*HQ,17000XXXXX,NBR,115117,310,26,02,1,1000,10,22,100119,FFFFFFFF#
```

Different commands can be detected according to the serial number (17000XXXXX). 2240.8116 corresponds to the latitude 22.408116, 11408.8108 at the latitude 11.4088108. There is no authentication, which allowed us to implement our own GPS client sending GPS coordinates. It can be seen Figure 4 that our GPS tracker is currently located in Pyongyang in North Korea as shown.

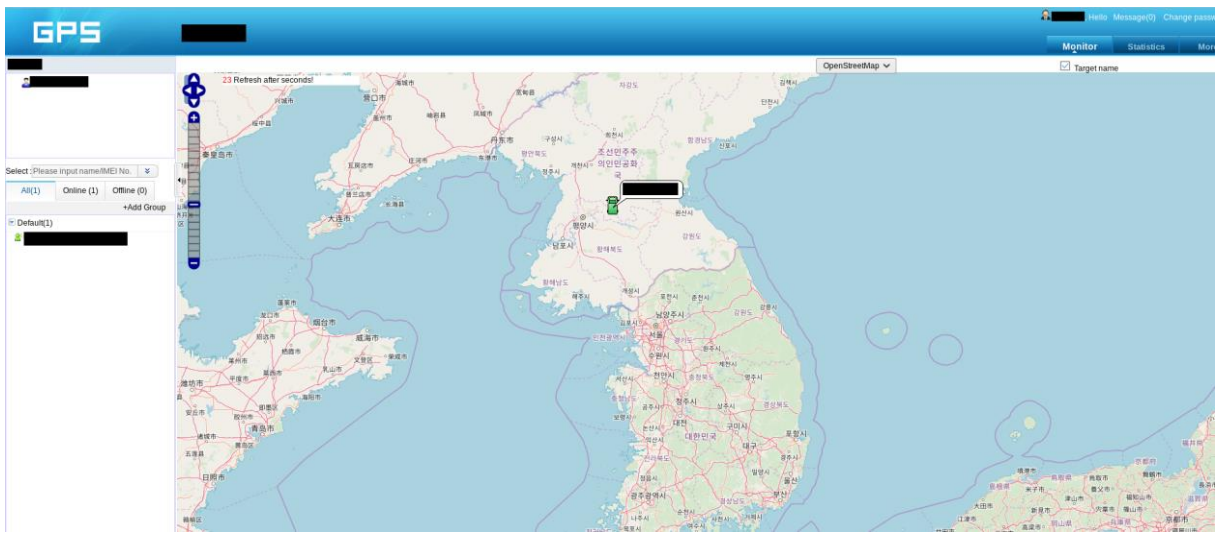

*Figure 4: Transmission of fake GPS coordinates through the vulnerable web interface without authentication*

A second (less expensive) tracker has no authentication to enable the server to verify the source of the collected data (through SMS or any data transmitted in 2G). The data sent to the management platform is always in the form:

```
yy.I.[S/N][BLOB-ASCII].~[BLOB-ASCII]
```

An interesting point is that it sends all SMS messages received on its radio interface to China, allowing the manufacturer to know exactly what the user is doing (network capture shown Figure 5 - analysis performed with Wireshark).

```
▶ Frame 316: 119 bytes on wire (952 bits), 119 bytes captured (952 bits)
  Raw packet data
▶ Internet Protocol Version 4, Src: 192.168.99.2, Dst: 203.130.62.29
▶ Transmission Control Protocol, Src Port: 58651, Dst Port: 8841, Seq: 118, Ack: 31, Len: 79
▼ Data (79 bytes)
    Data: 79790049f236393032313731323236313234363338393838...
    [Length: 79]

0000  45 00 00 77 6f 16 40 00  7f 06 5f 20 c0 a8 63 02   E··wo·@·  ··_ ··c·
0010  cb 82 3e 1d e5 1b 22 89  00 8e 28 10 3f e5 55 4d   ··>···"·  ··(·?·UM
0020  50 18 2a 80 12 b6 00 00  79 79 00 49 f2 36 39 30   P·*·····  yy·I·690
0030  32 31 37 31 32 32 36 31  32 34 36 33 38 39 38 38   21712261  24638988
0040  32 31 31 30 30 30 30 30  30 32 37 36 34 30 35 46   21100000  0276405F
0050  01 7e 31 39 30 31 30 39  31 30 35 34 31 37 0a 2b   ·~190109  105417·+
0060  34 34 30 30 32 35 32 33  39 01 06 53 74 61 74 75   44002523  9··Statu
0070  73 00 05 30 f3 0d 0a                               s··0···
```

*Figure 5: SMS forwarded to the manufacturer*

In response of sending an SMS "`Status`" to one of our GPS trackers, it has been observed that the targeted device sends a TCP packet to 203.130.62.29:8841/tcp, an IP geo-located in China, but actually located in the UAE and announced by the operator Etisalat, containing the message "Status", with the following additional information:

 690217122612463 corresponding to the S/N of the GPS tracker

 +440025239 being the number of the issuer of the SMS

This works for any SMS message, even if it does not correspond to a management SMS. An interesting use of this GPS Tracker would be to act as low-cost SMS relay system (<15 euros), by changing the management server IP address.

In general, the network security of GPS trackers is very low as it has been observed that the traffic is in plain text and is easily reverse engineered. Moreover, an attacker able to guess a serial number can send false information to the GPS management infrastructure.

### III.B. Firmware extraction and reverse engineering

The opening of the trackers made it possible to highlight the simplicity of the electronics used as we find mainly old MediaTek chipsets (MT6261 ARM) and GPS chipsets (eg U-BLOX). Debug interfaces are available for extracting firmware and default configuration items. Note that this part is not really a challenge since no protection against physical attacks has been integrated (Figure 5, opposite).

Once connected to the UART, the *mtd* blocks have been extracted. An analysis has been performed in order to detect potential hidden or undocumented commands. An example for one of the tested devices is shown in Figure 6. For this device, the embedded system is Nucleus RTOS and the OS size is typical for embedded hardware (4MB). Reversing of this dump allowed to reveal the presence of backdoors SMS codes in different GPS trackers. Unfortunately, not all of these SMS commands seem to be functional. This could be due to the use of the same firmware for different GPS trackers models of the same manufacturer.

The analysis did not reveal any major hidden features, but rather showed that the software seems to have been developed quickly due to the poor programming practices. It also has been observed that no firmware update features exist.

```
:OM:0006987F                DCB    0
:OM:00069880 aReboot         DCB    "*reboot*",0
:OM:00069889                DCB    0
:OM:0006988A                DCB    0
:OM:0006988B                DCB    0
:OM:0006988C                DCB    0
:OM:0006988D                DCB    0
:OM:0006988E                DCB    0
:OM:0006988F                DCB    0
:OM:00069890                DCB 0x2B ; +
:OM:00069891                DCB    0
:OM:00069892                DCB    0
:OM:00069893                DCB    0
:OM:00069894 a3646655        DCB    "*3646655*",0
:OM:0006989E                DCB    0
:OM:0006989F                DCB    0
:OM:000698A0                DCB    0
:OM:000698A1                DCB    0
:OM:000698A2                DCB    0
:OM:000698A3                DCB    0
:OM:000698A4                DCB 0x2C ; ,
:OM:000698A5                DCB    0
:OM:000698A6                DCB    0
:OM:000698A7                DCB    0
:OM:000698A8 aImeiset        DCB    "imeiset",0
:OM:000698B0                DCB    0
```

*Figure 6: Reverse engineering of the extracted firmware with IDA*

### III.C. Security of mobile applications

The manufacturer provides applications for iOS / Android smartphones, allowing the access to data gathered by the GPS trackers on a world map. Static and dynamic reverse engineering was performed on the Android application. The worst finding with the static analysis performed with the jadx utility [9] is an unsecured transfer of critical data over http.

```
<v:Envelope xmlns:i="http://www.w3.org/2001/XMLSchema-instance" xmlns:d="http://www.w3.org/2001/XMLSchema"
xmlns:c="http://schemas.xmlsoap.org/soap/encoding/" xmlns:v="http://schemas.xmlsoap.org/soap/envelope/"><v:Header
/><v:Body><GetTracking xmlns="http://tempuri.org/" id="o0" c:root="1"><DeviceID i:type="d:int">82383</
DeviceID><TimeZone i:type="d:string">UTC</TimeZone><MapType i:type="d:string">Google</MapType></GetTracking></
v:Body></v:Envelope>
HTTP/1.1 200 OK
Cache-Control: private, max-age=0
Content-Type: text/xml; charset=utf-8
Content-Encoding: gzip
Vary: Accept-Encoding
Server: Microsoft-IIS/7.5
X-AspNet-Version: 4.0.30319
X-Powered-By: ASP.NET
Date: Sun, 13 Jan 2019 07:17:09 GMT
Connection: close
Content-Length: 438

<?xml version="1.0" encoding="utf-8"?><soap:Envelope xmlns:soap="http://schemas.xmlsoap.org/soap/envelope/"
xmlns:xsi="http://www.w3.org/2001/XMLSchema-instance" xmlns:xsd="http://www.w3.org/2001/
XMLSchema"><soap:Body><GetTrackingResponse xmlns="http://
tempuri.org/"><GetTrackingResult>{"state":"0","deviceUtcDate":"2019-01-12
16:43:24","latitude":"39.056417","longitude":"126.257200","olatitude":"39.056417","olongitude":"126.257200","speed
":"0.00","course":"0","isStop":"1","icon":"27_0","distance":"0","acc":"0"}</GetTrackingResult></
GetTrackingResponse></soap:Body></soap:Envelope>
```

*Figure 7 - HTTP traffic analysis*

Further checking with dynamic analysis confirmed that the exchanges between the smartphone application and the remote server are done via http through a proprietary API system http://m.999gps.net/OpenAPIV2.asmx. We can see the identifier 82383 corresponding to the tracker (details depicted on Figure 7). It is also possible to access the Web Services Description Language (WSDL) of the API by adding WSDL to the address to retrieve a complete description of the service. Finally, the manufacturers provide an online debug interface as shown Figure 8.

*Figure 8- Debug interface of the API*

Spy the little Spies  10

By analyzing the traffic, it appeared that there is no authentication mechanism and that replaying the query parsed previously by Wireshark and changing the identifier makes it possible to retrieve the coordinates of other GPS trackers. This vulnerability was already reported by the Trackmaggedon team [1] in January 2018, but never been corrected by the manufacturer.

**III.D. Security of management websites and advanced attacks**

When buying a GPS tracker, it is provided with an identifier and a password for a website to access the information reported by the GPS tracker. By analyzing these websites, the following security flaws have been discovered:

- By default, the user name and password are the last 7 characters of the GPS tracker serial number. Users do not seem to know that they have to change the password and do not change passwords. It is thus possible for an attacker to provision all available accounts due to a lack of awareness
- It was possible, from the account of a GPS Tracker A to access to the coordinates of a GPS tracker B (in our possession) by specifying the ID of the tracker B in the http request of the tracker A - this is due to an insecure direct object reference vulnerability exploitable in the limit of having a random valid session. An example of the unauthenticated extraction of GPS coordinates history for a test tracker is depicted Figure 9.

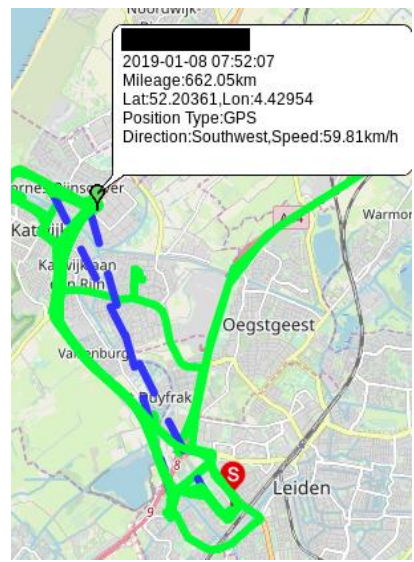

*Figure 9 - Un-authenticated extraction of the history of a tracker*

- Regarding, the specificity of GPS trackers inserted between the engine and the stop/start command with a remote control: it has been possible to add a geo-fence area to the trackers being used to remotely manage a vehicle engine, causing an alert if the vehicle leaves the area or allowing stop the engine stop. This has been successfully achieved through un-authenticated request send to the management interface targeting one of the test tracker as shown Figure 10.

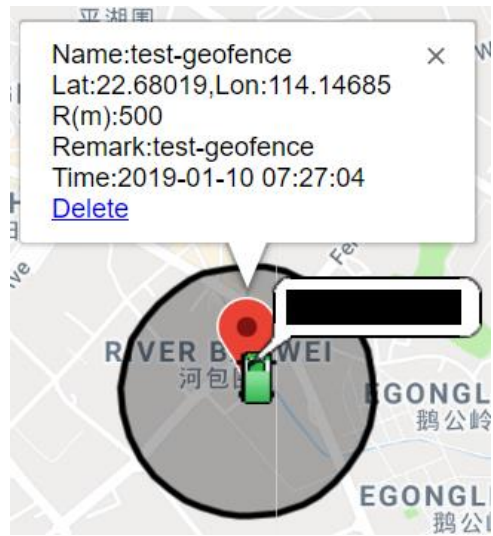

*Figure 10 – Un-authenticated Geo-fence added on a remote tracker*

- The interface provides access to the full history of GPS tracker data accessible through the same insecure direct object reference vulnerability.

## IV. Conclusion

As part of this study, we focused on the security and privacy issues of GPS trackers. It has been observed that various attack scenarios can be easily exploited in order to obtain the location information of the GPS trackers in an illegitimate way and that the protection of data exchanges was almost non-existent for all the tested devices. Configuration interfaces via SMS whose authentication is either non-existent or is easily circumvented via *CallerID* spoofing expose users to information leakage and to critical attacks due to the advanced features of certain models such as the remote control of vehicle engines. Moreover, many security issues have been found in APIs and websites as well as mobile applications.

Obviously, the different vulnerabilities are not limited to a single type of exploitation. The retrieved information was quite complete, including information from the surrounding telecommunication networks and GPS coordinates. These allows an attacker to effectively map the critical mobile telecom infrastructure of a given country. Trips and visited places are also

Spy the little Spies                                                                                                                   12

strategic information since, in correlation with work schedules, an attacker is able to know when and where a target works and where he lives. All the mentioned data is already present in the manufacturers servers recalling the software bug for the transfer of GPS coordinates of CelI-IDs in iOS [10].

## References

[1] V. Sykas, M. Gruhn, Multiple vulnerabilities in the online services of (GPS) location tracking devices, [online] https://0x0.li/trackmageddon/

[2] https://www.made-in-china.com/price/gps-jammer-price.html

[3] https://rntfnd.org/2017/09/15/gps-jammer-delays-flights-in-france/

[4] Kang Wang, Shuhua Chen, Aimin Pan, Time and Position Spoofing with Open Source Projects, BlackHat Europe 2015, Amsterdam, NL, [online] https://www.blackhat.com/docs/eu-15/materials/eu-15-Kang-Is-Your-Timespace-Safe-Time-And-Position-Spoofing-Opensourcely-wp.pdf

[5] https://github.com/osqzss/gps-sdr-sim

[6] 302 GPS TRACKER, [online] https://fccid.io/2AA64-302/User-Manual/User-Manual-3470390

[7] P. Barre "Kim", Multiple vulnerabilities found in Wireless IP Camera (P2P) WIFICAM cameras and vulnerabilities in custom http server, [online] https://pierrekim.github.io/blog/2017-03-08-camera-goahead-0day.html

[8] YateBTS GSM basestation - Open Source Software, [online] https://yatebts.com/open_source/#svn

[9] https://github.com/skylot/jadx

[10] https://www.apple.com/newsroom/2011/04/27Apple-Q-A-on-Location-Data/